 \documentclass[final,3p,times]{elsarticle}

\usepackage{amssymb}
\usepackage{lipsum}
\journal{Nuclear Physics A}

\begin{document}

\begin{frontmatter}



\title{Neural network-based prediction of particle-induced fission cross sections for r-process nucleosynthesis trained with dynamical reaction models}

\author[first,sec]{J.L.~Rodr\'iguez-S\'anchez\corref{cor1}}

\author[sec]{G.~Garc\'ia-Jim\'enez}

\author[sec]{H.~Alvarez-Pol}

\author[sec]{M.~Feijoo-Font\'an}

\author[first,sec]{A.~Graña-Gonz\'alez}

\cortext[cor1]{Corresponding author: j.l.rodriguez.sanchez@udc.es}

\affiliation[first]{organization={CITENI, Industrial Campus of Ferrol, University of Coruña},%Department and Organization 
            city={Ferrol},
            postcode={15403}, 
            country={Spain}}

\affiliation[sec]{organization={IGFAE, University of Santiago de Compostela}, 
            city={Santiago de Compostela},
            postcode={15482}, 
            country={Spain}}

\begin{abstract}
Large-scale computations of fission properties play a crucial role in nuclear reaction network calculations simulating rapid neutron-capture process (r-process) nucleosynthesis. Due to the large number of fissioning nuclei contributing to the r-process, a description of particle-induced fission reactions is computationally challenging. In this work, we use theoretical calculations based on the INCL+ABLA models to train neural networks (NN). The results for the prediction of proton-induced spallation reactions, in particular fission, utilizing a large variety of NN models across the hyper-parameter space are presented, which are relevant for r-process calculations.
\end{abstract}


\end{frontmatter}

%\tableofcontents



\section{Introduction}
\label{introduction}

Nearly half of the elements beyond iron are believed to be formed through the rapid neutron capture process, known as the r-process. However, the astrophysical sites of the r-process are not well known~\citep{Mumpower2016}. The proposed sites include neutrino driven winds in Type-II supernovae, neutron star (NS) or NS-Black hole mergers, collapsars, among others~\citep{Abbott2017,Siegel2019,Yong2021}. The detection of lanthanides in the red kilonova spectra of GW170817~\citep{Abbott2017} has provided clear evidence of r-process element production during NS merger events. In these scenarios, the decrease in nuclear stability in the r-process eventually ends when the heaviest nuclei become unstable to spontaneous fission. However, if the fission barrier is sufficiently low, neutron capture might induce fission instead of allowing the process to continue up the neutron drip line~\citep{Vassh2019}.

Nuclear fission, the process by which a heavy atomic nucleus divides typically into two lighter fragments, represents the clearest example of large-scale collective excitation in nuclei. The fission process is a unique tool for investigating the nuclear potential-energy landscape and its evolution as a complex function of excitation energy, elongation, mass asymmetry, and spin, as it progresses over the fission barrier and culminates at the scission point, resulting in the formation of fission fragments. The importance of fission in the r-process lies in its ability to recycle neutrons and produce a variety of heavy elements~\citep{Pinedo2007}. When fission occurs, it creates lighter neutron-rich nuclei that can undergo further neutron captures. This recycling process not only influences the abundance patterns of the heaviest elements but also helps to shape the final distribution of elements produced during the r-process, making fission a crucial factor in the synthesis of the Universe's heaviest elements~\citep{Eichler2015}. In particular, observables such as fission probabilities and the charge and mass distributions of fission fragments are critical in determining the final abundance patterns and refining our understanding of the r-process.

\section{Theoretical reaction models for neural network training}

The particle-nucleus collision is typically described using a two-step approach, commonly applied in spallation and charge-exchange reactions~\citep{Jcd2015}. The first step involves the collision itself, during which some nucleons from the target nucleus are either removed or modified. This process also results in the remnant gaining excitation energy and angular momentum. The second step consists of subsequent de-excitation processes, which occur through particle evaporation or, if applicable, through fission.

In this work, the collision between the light particle and the nucleus is modeled using the Liege Intranuclear Cascade model INCL++~\citep{Jason2020,jlrs2024}. The collision is represented as a sequence of binary interactions between nucleons (hadrons) within the system. Nucleons follow straight trajectories until they either collide with another nucleon or reach the surface, where they may escape. The latest version of INCL++ incorporates isospin- and energy-dependent nuclear potentials, calculated based on optical models, as well as isospin-dependent pion potentials. At the end of the collision, the nuclear system thermalizes forming a compound nucleus with some excitation energy and angular momentum.

The de-excitation of the compound nucleus is handled by the ABLA++ code~\citep{jlrs2022,JL2023}, which accounts for the emission of $\gamma$-rays, neutrons, light-charged particles, intermediate-mass fragments, or fission if the remnants are heavy nuclear systems. The fission decay width is initially described by the Bohr-Wheeler transition-state model, with subsequent corrections for dissipative and transient time effects through the solution of the Fokker-Planck equation. This method has yielded reasonable results for nuclei near the stability valley.

\section{Results}

The SoKAI NN framework~\citep{sokai} was trained using simulated data of known nuclei obtained from the coupling of INCL+ABLA models~\citep{jlrs2022,jlrs2024}, which provide accurate descriptions of experimental total reaction and fission cross sections. The input layer of the network received information on nuclear charge (Z), mass number (A), and the radii of proton (Rp) and neutron (Rn) density profiles. The network architecture consisted of four layers with a total of 184 trainable parameters. The activation functions used in the layers were LeakyReLU, Sigmoid, LeakyReLU, and LeakyReLU, respectively. Adam optimization was employed for back-propagation. The model's performance was evaluated using the absolute loss function, which showed that it reproduced both fission and total cross sections with an estimated error of 1-2\%. Notably, this model is capable of extrapolating results to exotic nuclei for which no experimental data or theoretical models are currently available.

\begin{figure}[b!]
	\centering 
	\includegraphics[width=0.48\textwidth]{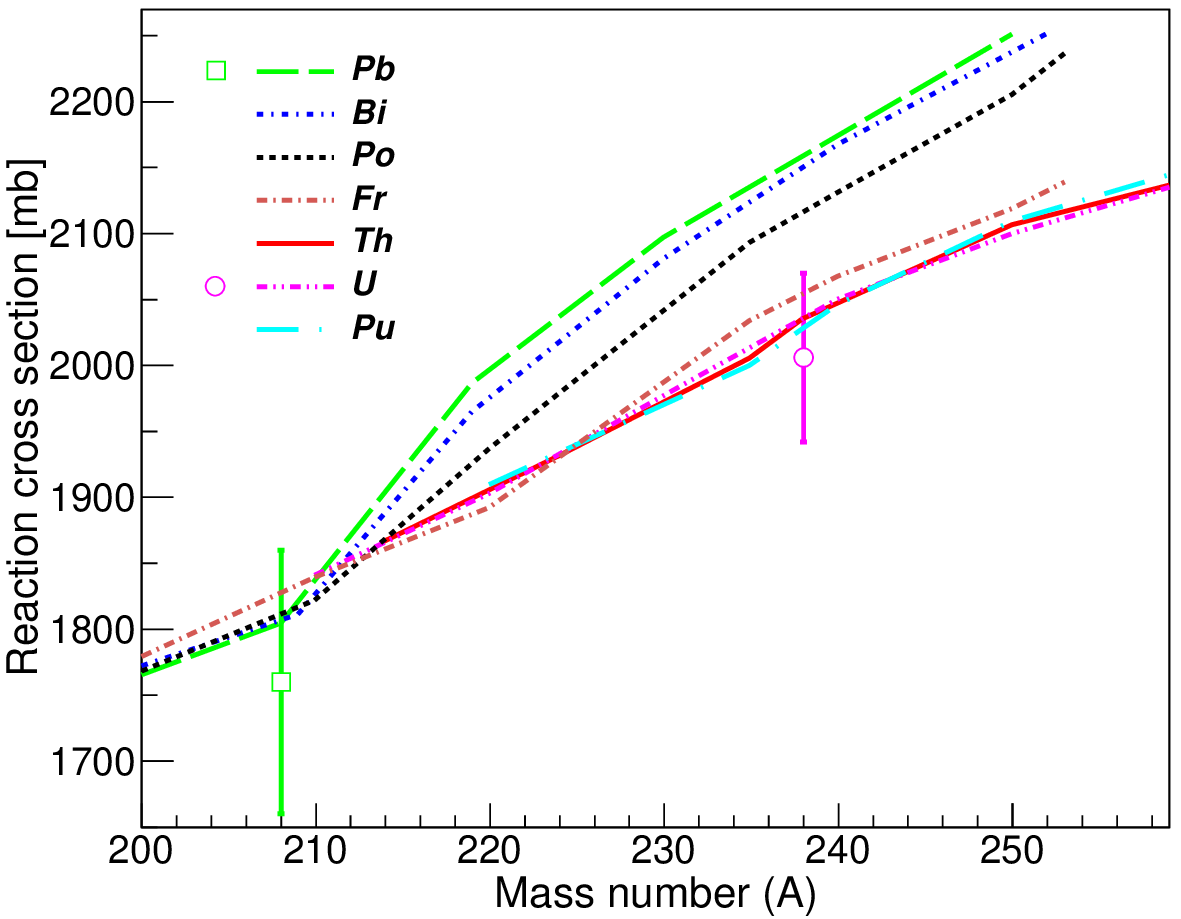}
	\includegraphics[width=0.48\textwidth]{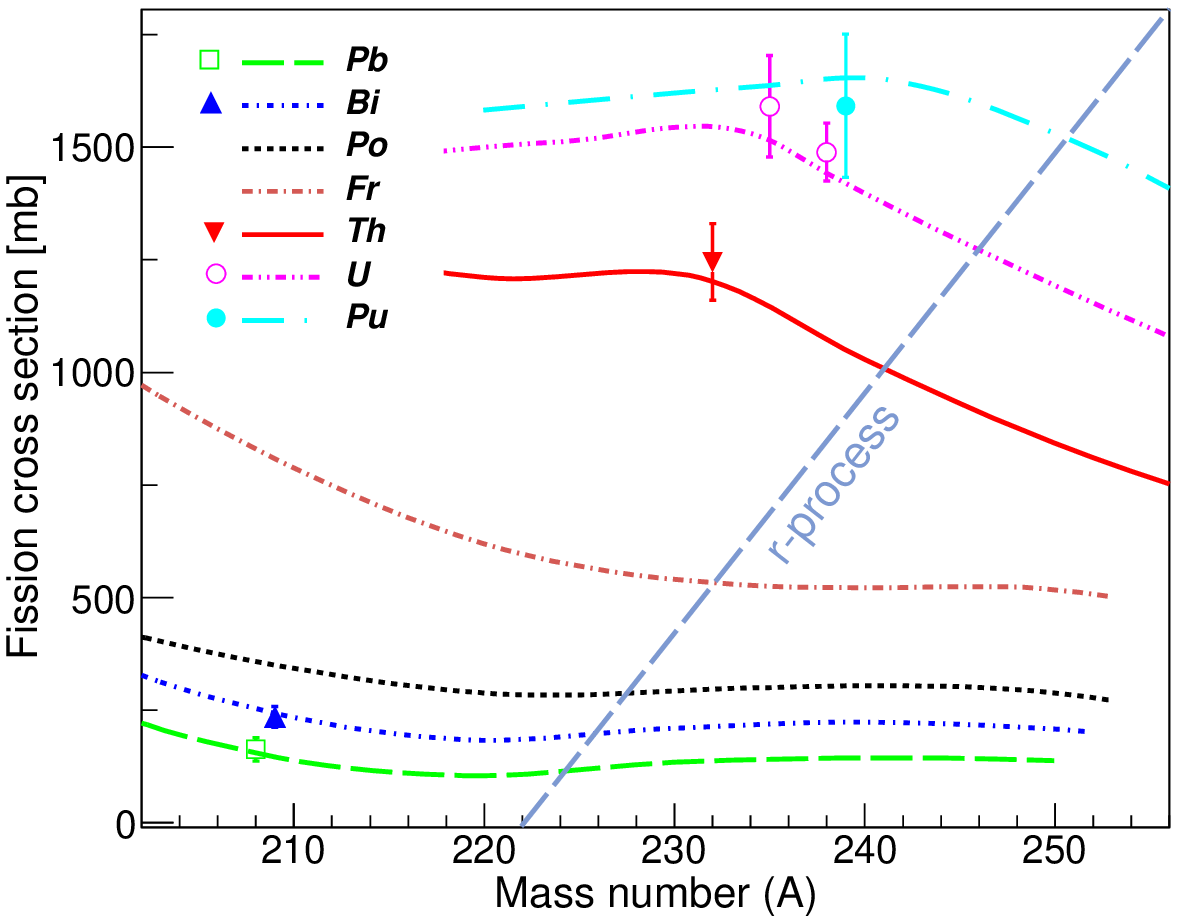}	
	\caption{(Left) Reaction cross section predictions for different elements as a function of the mass number A. (Right) Fission cross section predictions for different elements as a function of the mass number A. Symbols correspond to experimental data in both figures.} 
	\label{fig1}
\end{figure}

As shown in the Fig.~\ref{fig1}, our NN framework accurately predicts reaction cross sections for various stable isotopes and can also extrapolate to neutron-rich isotopes. Furthermore, it demonstrates strong agreement with experimental fission cross sections for stable isotopes and extends these predictions to isotopes beyond the currently known range.

While the current implementation of the SoKAI NN framework focuses on reproducing total reaction and fission cross sections, future developments will aim to extend this approach to accurately predict fission yields. This will be achieved by incorporating detailed information on the fission fragment mass and charge distributions before and after the emission of prompt neutrons, recently measured in inverse kinematics using transfer-, quasi-free nucleon knockout- and Coulomb-induced fission reactions~\cite{Ramos2018,Ramos2019,Ramos2020,Gabriel2023,Grana2023,Chatillon2019,Chatillon2022}. Additionally, the framework will be enhanced to include the multiplicity of prompt neutrons emitted after the scission point during the de-excitation processes, enabling a comprehensive description of the fission fragment distributions at all stages. This approach will not only improve the predictive power of fission yields but also allow for a more accurate estimation of their contributions to the elemental abundances observed in astrophysical environments~\cite{Vassh2019,Cowan2021}.

\section{Conclusions and perspectives}

The SoKAI NN framework has been trained with simulated data of known nuclei provided by the INCL+ABLA models. The NN architecture comprised four layers with a total of 184 trainable parameters. The training achieved a reasonable description of experimental data for total and fission cross sections of stable nuclei and effectively extends these predictions to isotopes beyond the currently known range given by the neutron-drip line. These promising results suggest that the NN could also be further trained with recent fission yield data of exotic nuclei measured at the GANIL and GSI/FAIR facilities through transfer-, quasi-free nucleon knockout-, and Coulomb-induced fission reactions in inverse kinematics~\citep{Ramos2020,Gabriel2023,Grana2023,Chatillon2022,Aumann2024}. Such an extension could enable accurate predictions of fission yields for very neutron-rich nuclei.

\section*{Acknowledgements}
We thank the support from the “María de Maeztu” Grants MDM-2016-0692 and CEX2023-001318-M, funded by MICIU/AEI /10.13039/501100011033. This work has been partly supported by the spanish funding agency for research (AEI) through projects PGC2018-099746-B-C21 and PID2021-125771NB-C21. J.L.R.-S. is thankful for the support by the Regional Government of Galicia under the program “Proyectos de excelencia” Grant No. ED431F-2023/43 and by the "Ram\'{o}n y Cajal" program under the Grant No. RYC2021-031989-I, funded by MCIN/AEI/10.13039/501100011033 and by “European Union NextGenerationEU/PRTR”.

\bibliographystyle{elsarticle-harv} 
%\bibliography{example}

\end{document}